\documentclass[aps,prb,twocolumn,showpacs,superscriptaddress]{revtex4-1}
\usepackage{graphicx}

\begin{document}

\title{Inelastic neutron scattering study of phonon density of states in nanostructured Si$_{1-x}$Ge$_x$ thermoelectrics}

\author{Chetan Dhital}
\affiliation{Department of Physics, Boston College, Chestnut Hill, Massachusetts 02467, USA}
\author{D. L. Abernathy}
\affiliation{Quantum Condensed Matter Division, Oak Ridge National Laboratory, Oak Ridge, Tennessee 37831-6393, USA}
\author{Gaohua Zhu}
\affiliation{Department of Physics, Boston College, Chestnut Hill, Massachusetts 02467, USA}
\author{Zhifeng Ren}
\affiliation{Department of Physics, Boston College, Chestnut Hill, Massachusetts 02467, USA}
\author{D. Broido}
\affiliation{Department of Physics, Boston College, Chestnut Hill, Massachusetts 02467, USA}
\author{Stephen D. Wilson}
\email{stephen.wilson@bc.edu}
\affiliation{Department of Physics, Boston College, Chestnut Hill, Massachusetts 02467, USA}

\begin{abstract}
Inelastic neutron scattering measurements are utilized to explore relative changes in the generalized phonon density of states of nanocrystalline Si$_{1-x}$Ge$_x$ thermoelectric materials prepared via ball milling and hot pressing techniques. Dynamic signatures of Ge-clustering can be inferred from the data by referencing the resulting spectra to a density functional theoretical model assuming homogeneous alloying via the virtual crystal approximation.  Comparisons are also presented between as-milled Si nanopowder and bulk, polycrystalline Si where a preferential low energy enhancement and lifetime broadening of the phonon density of states appear in the nanopowder. Negligible differences are however observed between the phonon spectra of bulk Si and hot pressed, nanostructured Si samples suggesting that changes to the single phonon dynamics above 4 meV play only a secondary role in the modified heat conduction of this compound.                  
\end{abstract}


\maketitle
\section{Introduction}
Recent advances in materials synthesis techniques have enabled grain boundary and strain engineering in a number of bulk, functional, nanocomposite alloys resulting in dramatic changes in their physical properties \cite{poudel, yu, renprl, luoptics, yang, hsu, zhao, dhital}.  One prominent example of this has been the ability to engineer nanostructured thermoelectric composites with substantially reduced thermal conductivities yet which retain reasonably robust conduction pathways, resulting in an enhancement in the thermoelectric figure of merit ZT \cite{poudel, yu, renprl, ma, dresselhouse}. While the enhanced efficiency of these thermoelectric nanocomposites continues to be optimized,\cite{yu} remarkably little is understood concerning the mechanism governing their thermal conductivity---the primary factor in their enhanced performance relative to their traditional bulk counterparts.

One of the simplest and most striking examples of an enhanced ZT in a bulk theromoelectric nanocomposite was recently reported in a series of Si$_{1-x}$Ge$_x$ alloys prepared via high energy ball milling and hot pressing techniques \cite{renprl}.  A substantial drop in the thermal conductivity of P-doped, nanostructured, Si can be achieved relative to bulk Si, and this gain is further extended when additional point defect scatterers are introduced via Ge-alloying.  The resulting Si$_{1-x}$Ge$_x$ nanocomposite material allows both grain boundary and point defect phonon scattering mechanisms to be tuned while optimizing the efficiency of the resulting fused nanocrystalline network.  The effect that these added scattering mechanisms have in the distribution of the phonon density of states (PDOS) of the resulting nanocomposites however remains an ongoing area of interest\cite{sopu}.

Modifications to the PDOS in nanocrystalline alloys due to both finite-size and grain boundary effects have been previously explored via a number of probes where phonon lifetime broadening effects\cite{fultzphilmag} as well as both red-\cite{yangraman, gupta} and blue-shifts\cite{wangraman} in phonon frequencies have been reported.  Additionally, strain fields introduced via the crystallite reduction process and frozen at grain boundaries can modify the resulting PDOS through renormalized force constants due to bond length changes near interfaces \cite{kara}.  Enhancements in the low energy spectral weight of the PDOS in a number of systems are often ascribed as interface modes or intercrystallite vibrations, and lifetime broadening effects are typically observed to reduce the sound velocity in the low energy acoustic regime of nanocrystallites \cite{fultzreview, fultzjap, frase}.  The interplay between these and confinement effects associated within the lattice dynamics of nanocrystals however remains an active area of investigation where experimental studies mapping the phonon mode distributions in nanocomposites with anomalous thermal transport properties are rather scarce. 

In this article, we report inelastic neutron scattering measurements mapping the phonon density of states across a series of nanocomposite Si$_{1-x}$Ge$_x$ ($x=$ 0, 0.05, 0.10, 0.20) samples clustered near the Si-rich end of the phase diagram that were previously shown to possess substantially reduced thermal conductivity\cite{renprl}.  The effect of adding point defect Ge scattering sites (in the low concentration limit) into the hot pressed nanocrystalline matrix as well as modifications to the PDOS imparted via the nanotexturing process are explored. At higher Ge concentrations, the shift in spectral weight observed upon the introduction of Ge into the Si$_{1-x}$Ge$_x$ matrix deviates from expectations of theoretical estimates assuming a homogeneous Si-Ge mixing that smoothly renormalizes the lattice force constants.  Instead, signatures of a two-mode behavior and Ge clustering appear in the resulting spectra\cite{fenren} as Ge moves far away from the point-defect regime. In comparing the PDOS of bulk, polycrystalline Si with hot pressed, nanostructured Si (whose thermal conductivity has been reduced by nearly an order of magnitude), the phonon spectra of the two compounds appear nearly identical.  This suggests a minimal role of modified single phonon dynamics above our experimental resolution in changing the heat conduction mechanism of this thermoelectric system.    

Our results also allow a comparison between bulk polycrystalline Si and as-milled Si nanopowder at three different temperatures (300 K, 500 K, and 700 K).  While the PDOS of bulk Si is nearly temperature independent up to 650 K, an anomalous enhancement in the lattice dynamics of the nanopowder Si appears near 35 meV upon warming.  At 300 K, a lifetime broadening of phonon modes in the transverse acoustic energy regime appears in the nanopowder spectrum, resulting in an enhancement of the low energy PDOS relative to the bulk. More generally, the competition between energy scales in nanocomposites such as energy raising surface effects versus free energy lowering enhanced entropies intrinsic to the nanostructures themselves (such as vibrational entropy, $S_{vib}$\cite{fultzreview}) remains unexplored in this material.  Our measurements allow the calculation of the vibrational entropy change of Si upon transitioning from the bulk crystalline phase to the highly strained ball milled nanopowder phase and show a sizeable increase in $S_{vib}$.  

\section{Theoretical Model of PDOS}
As a guide for interpreting our experimental results, the one-phonon density of states (DOS), $g(\omega)$, is caluclated as:  $g(\omega)=\frac{1}{(2\pi)^3}\sum_j\int\partial(\omega-\omega_j(\textbf{q}))\partial\textbf{q}$  where $\omega_j(\textbf{q})$ is the frequency of a phonon mode in branch $j$ with wave vector, $\textbf{q}$, and the integral, taken over the first Brillouin zone, is evaluated using the tetrahedron approach.  The phonon frequencies are obtained by diagonalizing the dynamical matrix.  The harmonic interatomic force constants for Si and Ge are calculated from first principles within the framework of density functional perturbation theory (DFT).  Here we have used the Quantum Espresso program \cite{baroni}. An $8\times 8\times 8$ Monkhorst-Pack \cite{monkhorst} mesh has been used for the electronic states, and a cut-off energy of 30 Ry was chosen. The pseudopotentials were generated following the method of von Barth and Car \cite{barth}. These pseudopotentials accurately reproduce the measured phonon dispersion data for both Si and Ge. The exchange-correlation energy is calculated within the local density approximation, using the results of Ceperley and Alder \cite{ceperly} as parameterized by Perdew and Zunger \cite{perdew}. For the Si$_{1-x}$Ge$_x$ alloy, we have adopted a virtual crystal approximation model, where the interatomic force constants and atomic masses of pure Si and Ge crystals are averaged according to their relative concentrations in the alloy.  Further details of this approach have been previously described elsewhere \cite{broido, ward, kundu}.

\section{Experimental Details} 

Three nanocomposite Si$_{1-x}$Ge$_x$ samples with $x=$0.05, 0.10, 0.20 were prepared via high energy ball milling and hot pressing techniques identical to the processes described earlier \cite{yu,renprl}. All nanopowder samples were ground in a SPEX 8000 high-energy ball mill inside of stainless steel jars. These samples were then hotpressed inside a custom built press with DC graphite heating elements and densified as described earlier \cite{yu}.  We note here that each of these hot pressed concentrations had $2\%$ P added to the matrix to enable sample compaction and increase the carrier concentraion.  Each alloy was subsequently verified to be single phase via x-ray diffraction measurements, and EDS measurements confirmed the absense of contaminants from the ball milling process to within 1$\%$.  We note here that hotpressed samples prepared in this way are very stable.  Previous measurements failed to observe significant property and structure changes following sample annealing at $1100$ C in air for over one month.  

Our study also looked at three Si samples:  (1) a bulk Si sample consisting of mortar crushed (5N) Si ingots, (2) a nanopowder Si sample created by ball milling the same bulk Si ingots in an argon atmosphere, and (3) a hotpressed Si sample consisting of Si and $2\%$P ball milled together and then hotpressed in a manner identical to the Si$_{1-x}$Ge$_x$ nanocomposite samples.  For the remainder of the paper's discussion, we refer to the ball milled and hot pressed SiP$_{0.02}$ sample as ``Hot pressed Si'', the bulk mortar crushed polycrystalline Si ingots as ``Bulk Si'', and the loose ball milled Si nanopowder as ``Nanopowder Si''. All samples were stored under an argon atmosphere and eventually loaded into sealed aluminum cans within a furnace for neutron measurements.  

X-ray measurements were collected in a Bruker D2-Phaser system, and crystallite imaging was performed on a JEOL 6340F scanning electron microscope (SEM).  Neutron measurements were performed on the ARCS time-of-flight chopper spectrometer at the Spallation Neutron Source at Oak Ridge National Lab\cite{abernathy}.  A series of different incident energies and chopper frequencies were utilized to explore the lattice dynamics with variable instrument resolution across different energy ranges.  The two configurations chosen to map the phonon density of states were $E_i=120$ meV and $E_i=40$ meV  with incident Fermi chopper frequencies of $600$ Hz and $300$ Hz respectively.  Detector efficiencies were corrected via a standard vanadium normalization, and the background scattering for a given incident energy and temperature was removed by subtracting off from an empty sample can under the sample experimental conditions.  Samples were loaded into cylindrical $1/2$ inch inner diameter Al cans inside a He atmosphere and sealed prior to measurement.   

The background subtracted data was summed over all $\textbf{Q}=0 - 14 \AA^{-1}$ ($E_i=120$ meV) and $\textbf{Q}=0 - 8 \AA^{-1}$ ($E_i=40$ meV) for neutron energy loss spectra.  Data neighboring detector gaps or edges in (Q,E)-space were excluded from the summations in order to minimize detector noise effects within the spectra.  Even though Si possesses a strong coherent scattering cross section, the broad momentum averaging of intensities over multiple Brillouin zones generates a reasonable approximation of the incoherent phonon density of states, which is verified in Section IV of this paper. 

The integrated intensity spectra were then fit via a routine similar to that detailed by Kresch \textit{et al.}\cite{kresch}: (1) removing an initial, constant, multiphonon sum from the scattering data, (2) calculating the resulting single-phonon scattering profile, (3) determining the corresponding multiphonon contribution up to 4-phonon scattering processes by convolving the single-phonon scattering profile, (4) scaling this multiphonon contribution upward to approximate multiple scattering contributions, and (5) iteratively repeating this process until convergence was reached.  The multiple scattering contribution (which is geometry and packing density dependent) was typically of the order unity with the multiphonon contribution and varies no more than $\approx7\%$ over the range of all samples.  For $E_i=120$ meV data, a small constant background term had to be added to the data due to fast neutron/gamma background scattering from the neutron source.  For all samples, a hard single-phonon cutoff frequency of $66$ meV was enforced, which is supported by the raw scattering data.  

In order to compare data of different $E_i$'s and resolution profiles, density of states data from $E_i=120$ meV and $E_i=40$ meV were normalized together at $\Delta E=28$ meV (near a minimum in the PDOS of Si).  Specifically, in combining the data from each $E_i$, the total spectral weight of the single phonon spectrum in the $E_i=40$ meV data was normalized to the percentage of the density of states below 28 meV in the $E_i=120$ meV spectra.  Combing data in this fashion necessitates the approximation that the multiphonon background can be smoothly scaled; however we stress here that our analysis holds for comparing each data set individually as well as combined.  All relative changes reported between each density of states spectra are determined within an individual $E_i$ configuration, and we cross-checked the normalization procedure through comparison of the bulk Si PDOS with the DFT model (discussed further in Section IV).  

The resulting generalized phonon density of states (GPDOS) is a neutron weighted average over the atomic species, $d$, where the GDPOS $\propto\sum_d  g_d(E)\gamma_d\frac{\sigma_d}{m_d}$ (neglecting the variation in Debye-Waller factors).  Here, $g_d(E)$ is the partial density of states for atom $d$, $\gamma_d$ is the fraction of atomic occupation, $\sigma_d$ is the total neutron scattering cross-section and $m_d$ is the mass of the $d^{th}$ atomic species.  For pure Si samples studies, the GPDOS is identical to the conventional PDOS.  There is an approximately $35\%$ decrease in this weighting factor when comparing Ge to Si; however the GPDOS derived from our measurements of Si$_{1-x}$Ge$_x$ samples with $x\leq0.20$ remain largely dominated by the Si scattering. 

\begin{figure}
\includegraphics[scale=.45]{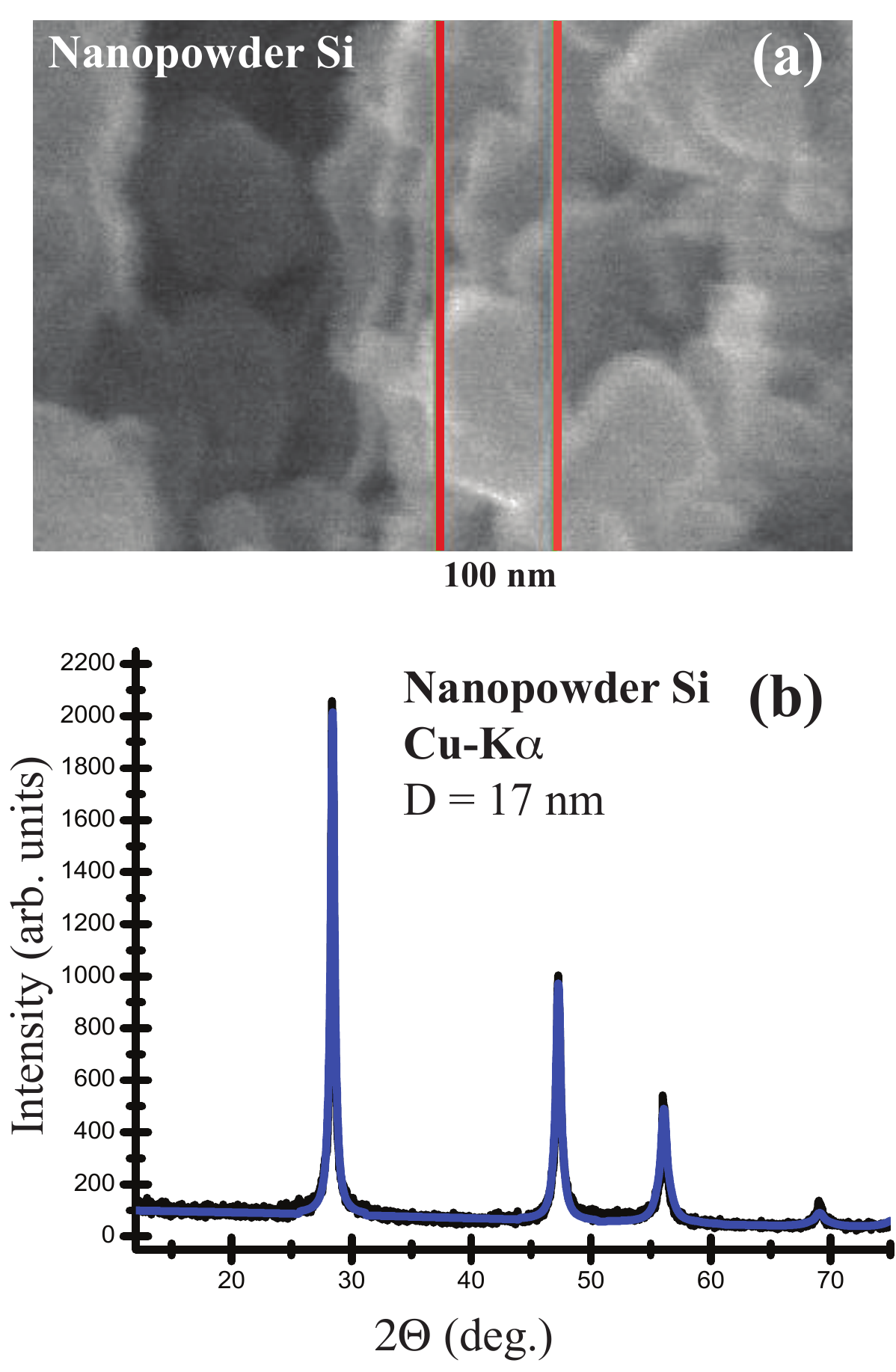}
\caption{(a) SEM image of a portion of the Si nanopowder sample used in our studies.  (b) X-ray diffraction profile from the same Si nanopowder sample showing coherent scattering domains of approximately 17 nm in diameter.}
\end{figure}

\begin{figure}
\includegraphics[scale=.325]{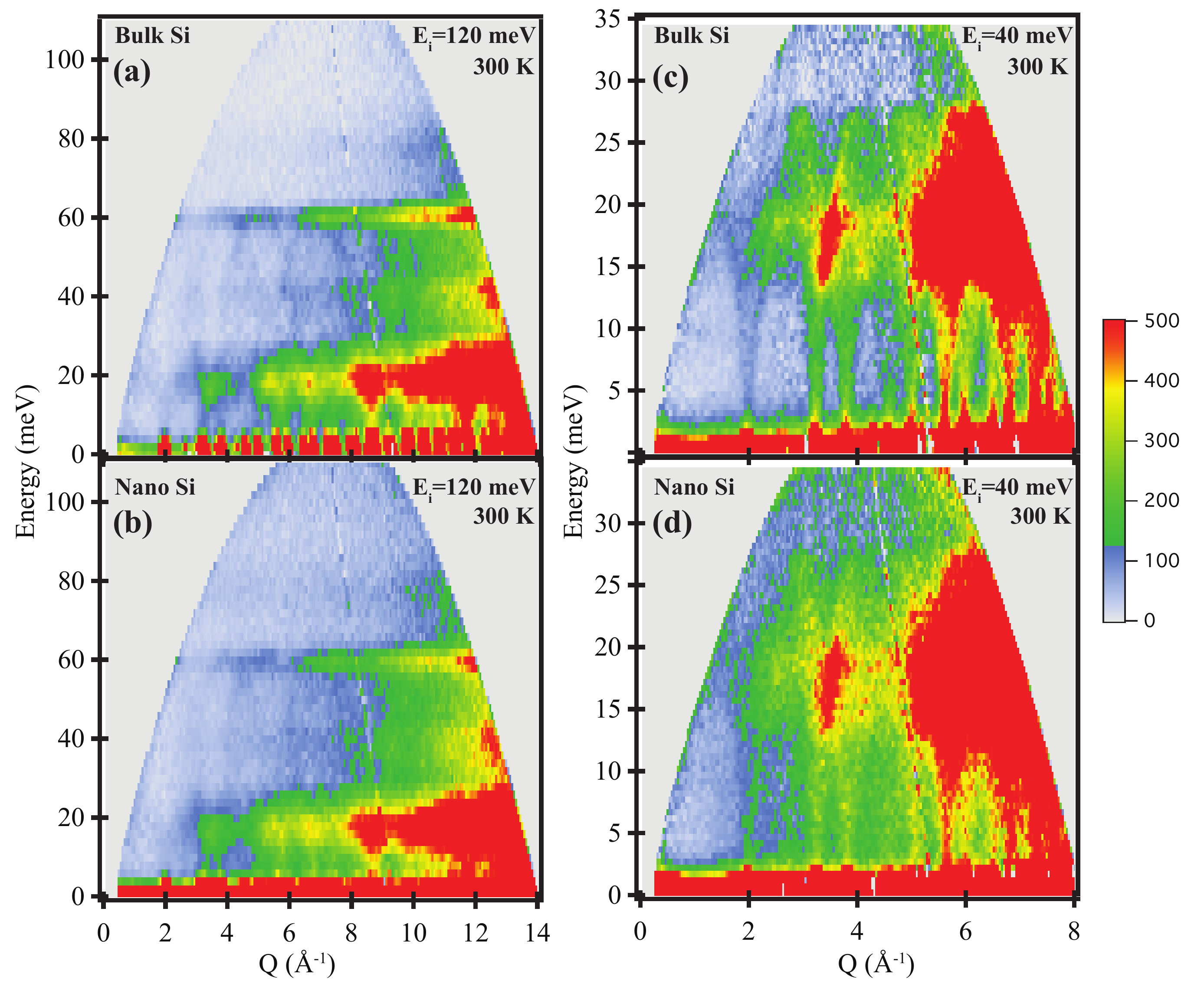}
\caption{(\textbf{Q}, $\omega$) intensity maps showing background subtracted inelastic neutron scattering spectra with incident neutron energies $E_i=120$ meV for both the (a) bulk Si and (b) nanopowder Si samples.  Intensity maps with $E_i=40$ meV are also shown in panels (c) and (d) for the bulk and nanopowder Si samples respectively.}   
\end{figure}

\section{GPDOS in bulk crystalline S\lowercase{i} and nanopowder S\lowercase{i}}

As an initial investigation into the effect of high-energy ball milling bulk crystalline Si into a nanopowder, both the bulk Si powder and ball milled Si nanopowder were explored.  SEM images in Fig. 1 (a) show the resulting particle size of the Si nanopowder to be $\approx100$ nm; however the x-ray scattering profile shown in Fig. 1 (b) shows a much smaller correlation length due to a high defect density and enhanced strain field imparted via the ball milling process.  Fits to a Williamson-Hall style plot \cite{williamson,dhital} render a coherent diffraction domain size of $\approx 17$ nm---suggesting a large density of grains within the mean particle size resolved via SEM images.  Within resolution, no volume fraction of amorphous Si was observed in diffraction measurements of this or any of the other nanotextured samples studied in this paper.  

Looking first at the comparative lattice dynamics at $T=300$ K, the Al-can background subtracted intensity maps of both the bulk polycrystalline and nanopowder Si samples are plotted in Fig. 2.  From the $E_i=120$ meV data, both samples show a well defined single phonon cutoff energy of $66$ meV above the transverse optical zone boundary and Van Hove singularity\cite{nilsson, nelin} with a broad multiphonon peak appearing at $\approx78$ meV.  From an initial inspection of both the $E_i=120$ meV and $E_i=40$ meV maps, an overall broadening of spectral features in both momentum and energy appears in the nanopowder sample relative to the bulk---consistent with the shortened correlation length and lifetime effects of excitations expected in a highly disordered lattice.  In order to quantify the resulting effect on the GPDOS, the scattering for both samples was integrated over momentum and plotted in Fig. 3.

After following the iterative fitting procedure for the removal of multiphonon and multiple scattering effects detailed in Section II, the resulting single-phonon and scaled multiphonon components of the total scattering are overplotted with the momentum integrated data in Fig. 3.  The solid line shows the combined scattering terms and the resulting fit to the total integrated scattering.  The refinement of the multiphonon scattering components is governed by scattering above the single-phonon cutoff where a pronounced multiphonon peak in the momentum integrated data appears in all data sets.  This provides a fortuitous guideline for removing multiphonon contamination, and the resulting fits to the data model the $\Delta E>66$ meV regime and the multiphonon peak near 78 meV well.  

The only exception to this is a high frequency upturn and an enhanced constant background scattering term observed only in the nanopowder Si sample at 300 K.  While the enhanced energy-independent scattering term can be removed, our multiphonon fits fail to model the slight intensity upturn observed for $\Delta E>100$ meV.  Both of these spurious high frequency features however disappear upon heating the Si nanopowder sample, suggesting either the influence of an unknown contaminant adsorbed to the nanopowder (despite the precaution of maintaining an Ar environment during preparation and storage) or the influence of anomalous grain boundary/strain effects in the phonon spectrum.

\begin{figure}
\includegraphics[scale=.35]{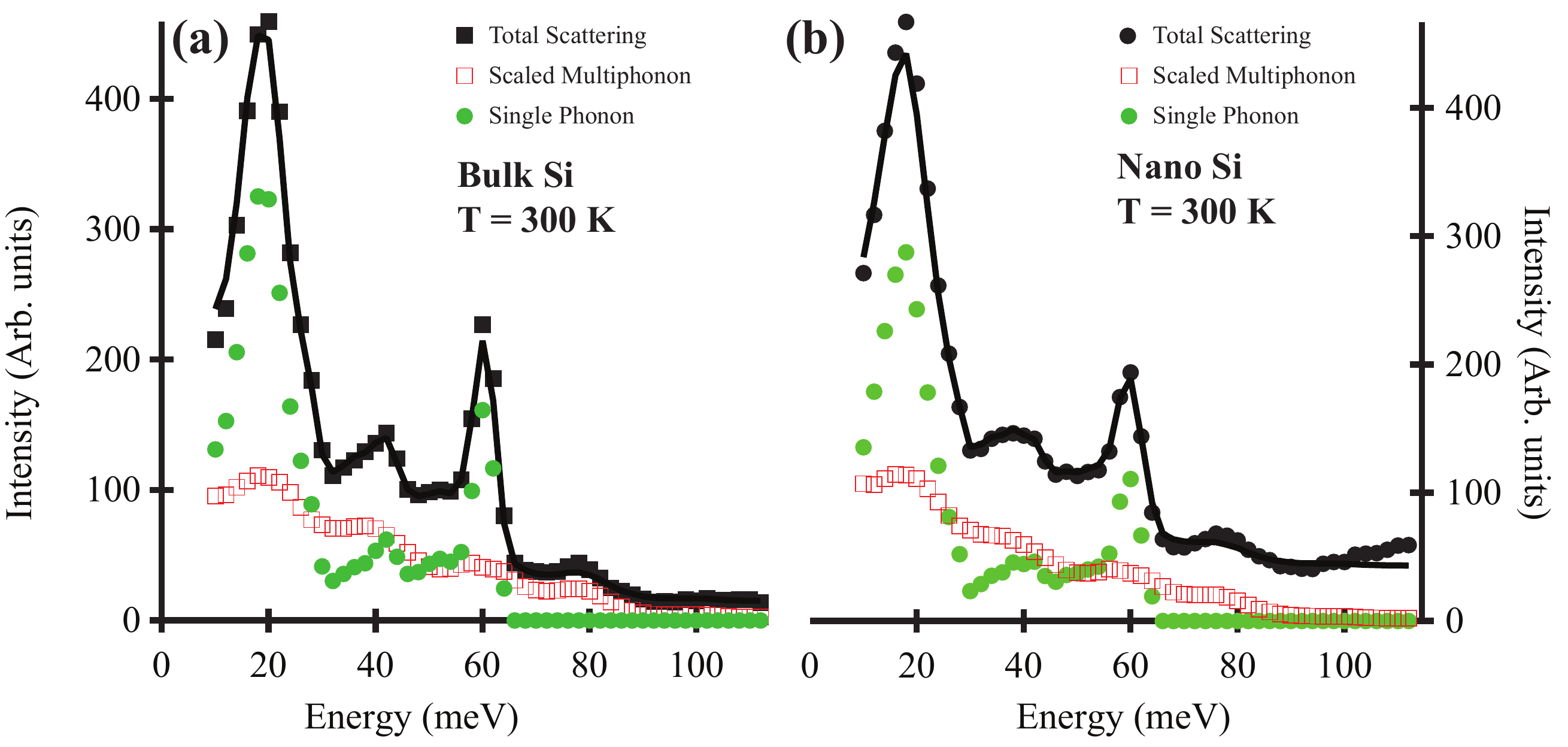}
\caption{Momentum integrated and averaged intensities (black squares) for both (a) bulk Si and (b) nanopowder Si samples.  Closed circles show the reduced single-phonon component of the scattering, and open squares show the multiphonon/multiple scattering component of the total intensities.  Solid lines show the combined single phonon, scaled multiphonon, and energy independent background intensities fit to the total scattering data.}
\end{figure}
 
\begin{figure}
\includegraphics[scale=.45]{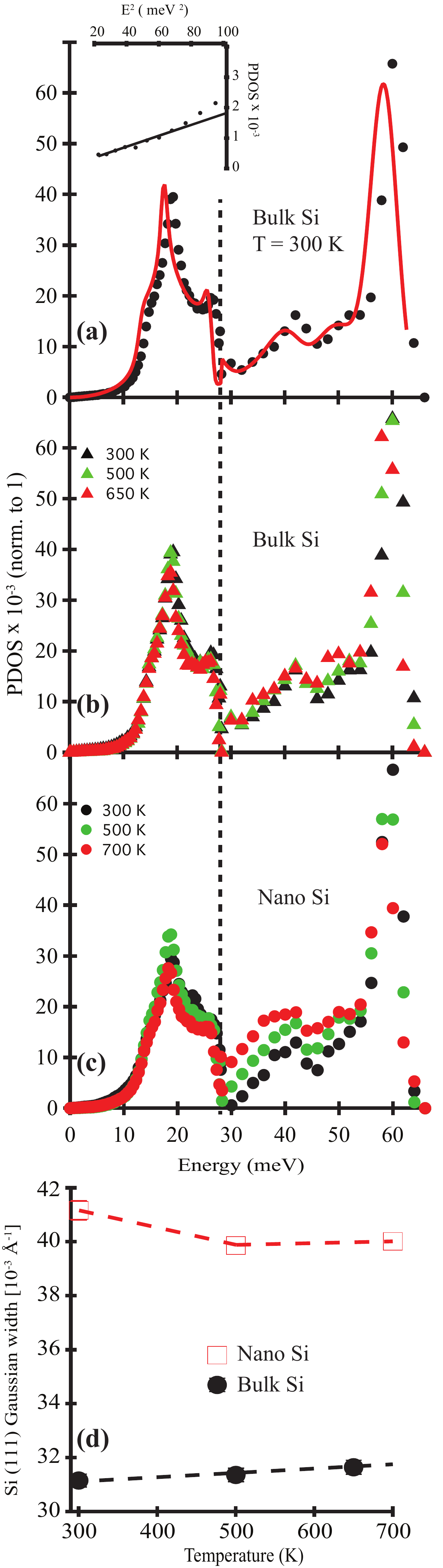}
\caption{(a) Black circles show the reduced PDOS for bulk Si at 300 K.  The red solid line shows the DFT calculated PDOS for Si as described in the text.  Inset shows the low energy PDOS as a function of $E^2$ where symbols show the PDOS of Si and the solid black line shows a linear fit to the low energy PDOS. (b) Reduced PDOS for bulk Si are compared at 300 K, 500K, and 650 K.  (c) The reduced PDOS for nanopowder Si are compared at 300K, 500K, and 700 K.  Dashed line denotes the energy crossover between $E_i=120$ meV and $E_i=40$ meV data. (d) Q=(1,1,1) Bragg peak widths plotted as a function of sample temperature for the nanopowder and bulk Si samples (errors within symbols).}
\end{figure}

The resulting PDOS obtained from the single-phonon profile from the bulk Si sample is plotted in Fig. 4 (a).  As described in Sect. II, data from the $E_i=120$ meV dataset transitions to the $E_i=40$ meV dataset below $\Delta E=28$ meV (denoted by a dashed line) in order to utilize the superior resolution afforded by a lower incident energy over the acoustic phonon regime.  As a reference we have overplotted the results of our DFT calculations (convolved with the instrument resolution) for the PDOS of Si with the experimentally determined PDOS.  Aside from a small frequency shift due to the systematic underestimation of zone boundary energies in the DFT calculation, the experimental PDOS matches remarkably well with the theoretical expectation.  This provides an added degree of confidence in the experimental data reduction and normalization procedure as well as the momentum summation's incoherent approximation to the PDOS.  As a further check, the low energy PDOS is plotted versus $E^{2}$ in the inset of Fig. 4(a).  In the long wavelength limit, the PDOS$\propto E^{2}$ as expected; however this approximation breaks down quickly above $\Delta E=9$ meV in the experimental data and above $\Delta E\approx5$ meV in the DFT simulation.

\begin{figure}
\includegraphics[scale=.55]{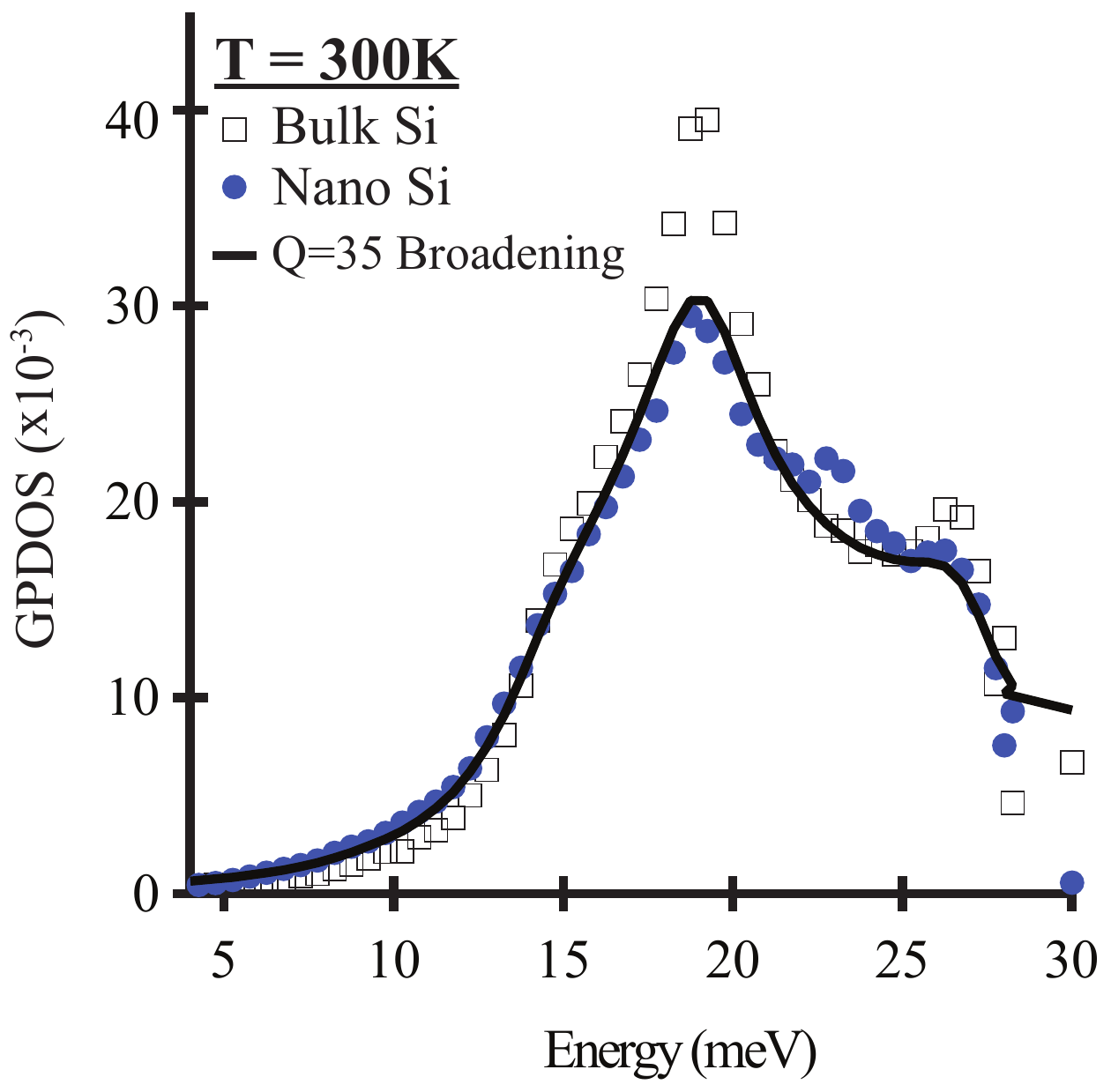}
\caption{Expanded comparison of the 300 K PDOS of bulk (open squares) and nanopowder (solid circles) Si below 30 meV.  The solid line shows the measured PDOS of bulk Si convolved with a damped harmonic oscillator lineshape as described in the text.}
\end{figure}

\begin{figure}
\includegraphics[scale=.5]{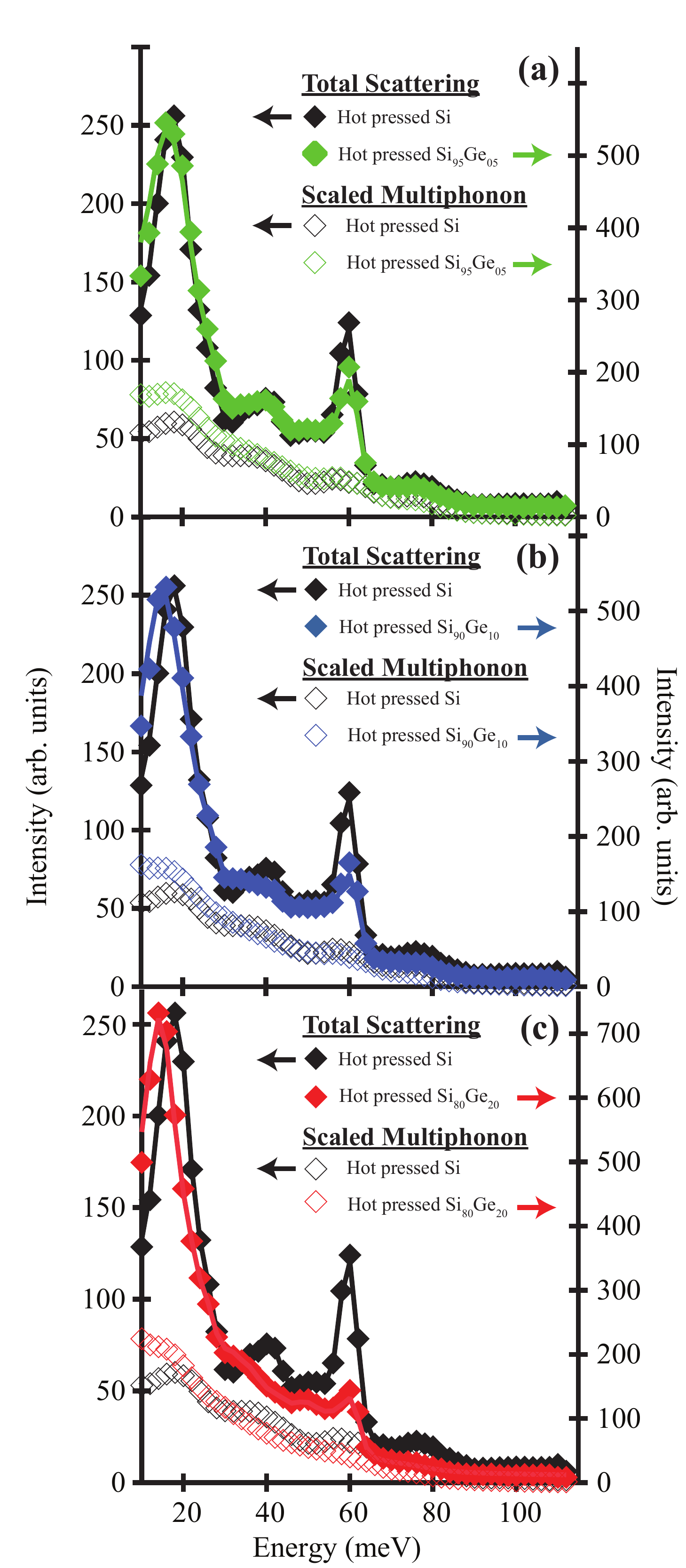}
\caption{$E_i=120$ meV data showing the resulting single phonon, scaled multiphonon, and total scattering fits overplotted with the momentum averaged intensity data for (a) Si$_{0.95}$Ge$_{0.05}$, (b) Si$_{0.90}$Ge$_{0.10}$, and (c) Si$_{0.80}$Ge$_{0.20}$.  As a reference, overplotted in each panel are the corresponding single phonon, scaled multiphonon, and momentum averaged total scattering intensities for the Ge-free hot pressed Si sample.}
\end{figure}

Now applying an identical reduction procedure to both bulk and nanopowder Si samples at higher temperatures, the temperature dependence of the PDOS for each sample at $T=$ 300 K, 500 K and $700$ K are plotted in Figs. 4 (b) and (c).  As expected, the resulting PDOS for bulk Si remains largely unchanged upon heating, although a slight anharmonicity and redshift of the optical zone boundary zone boundary Van Hove feature appears at $650$ K.  In contrast, the nanopowder Si sample shows more dramatic changes in the PDOS upon heating. At increased temperatures, an anharmonic redshift again appears in the optical zone boundary feature of the nanopowder; however there also appears a redistribution of spectral weight resulting in an enhanced PDOS for $30\leq E\leq 50$ meV.  The PDOS of the nanopowder sample in the acoustic regime, while broadened at 300 K, sharpens slightly upon heating likely due to grain growth and annealed strain effects. 

We note here that the sample's temperature history went from 300 K to 700 K and finally back down to 500 K. This suggests that the increased sharpening of the acoustic mode distribution from the 700 K and 500 K data is likely due to the continued annealing after the 700 K data collection ($\approx4$ hrs).  As a rough check, the peak widths of the Si Q=(1,1,1) Bragg reflections were analyzed for both the nanopowder and bulk Si samples as a function of temperature and plotted in Fig. 4 (d). A slight narrowing of the nanopowder Bragg widths was observed upon heating to 700 K which supports the notion of the acoustic mode lifetime sharpening as arising from an annealing/grain growth effect.  The peak widths however remain notably broader than those of bulk Si (the spectrometer's resolution) after heating, demonstrating that the sample remains in a largely nanotextured phase.  

The spectral weight enhancement above $30$ meV in the nanopowder however is robust and scales monotonically with temperature, suggesting an origin intrinsic to the nanostructured sample.  This contrasts with the naive expectation of an annealing effect such as that observed at lower energies which should smoothly transition the dynamics from the nanopowder to those of the bulk.  At present, the origin of this enhancement is unclear, although grain boundary modes or defects may play a role.  Specifically, these modes can in principle be unpinned and thermally activated upon heating and would renormalize the spectral weight of the PDOS distribution if the volume fraction of defects was large enough.                  

One striking feature in comparing the PDOS of the nanopowder and bulk Si samples is the preferential broadening of the spectral weight distribution in the transverse acoustic regime.  There is little difference in the spectral shape near the Van Hove singularities in the optical regime of the PDOS, but a clear phonon broadening is resolvable below 30 meV. The broadening at low energies is highlighted in a comparison of the bulk and nanopowder samples below 30 meV plotted in Fig. 5. In order to estimate the magnitude of broadening, a conventional method of modeling the spectrum is to convolve the ideal spectrum with a damped harmonic oscillator function \cite{delaire, fultz1997}. The most practical approach is then to first consider the measured bulk Si phonon PDOS as a model starting point.  

We have already shown that the calculated PDOS convolved with the instrument resolution matches the measured spectrum for bulk Si well. Therefore, we can assume the intrinsic line widths of the bulk Si spectrum are well below the instrument resolution, and we can convolve the function $\Gamma (E)\propto\frac{1}{(E_0/E-E/E_0)^2+(1/\beta)^2}$ \cite{fultzreview, delaire} with the experimentally obtained Si PDOS to determine the needed $\beta$ to approximate the nanopowder response.  The best fit obtained for the nanopowder yields a resulting $\beta=35$.  The quality factor, $\beta$ is proportional to the number of wavelengths a phonon of a given frequency is able to propagate within a crystallite domain, and the relatively high value for our Si nanopowder can be attributed to the larger Si crystallite sizes relative to those in earlier studies \cite{fultz1997, frase}.     
    
\begin{figure}
\includegraphics[scale=.5]{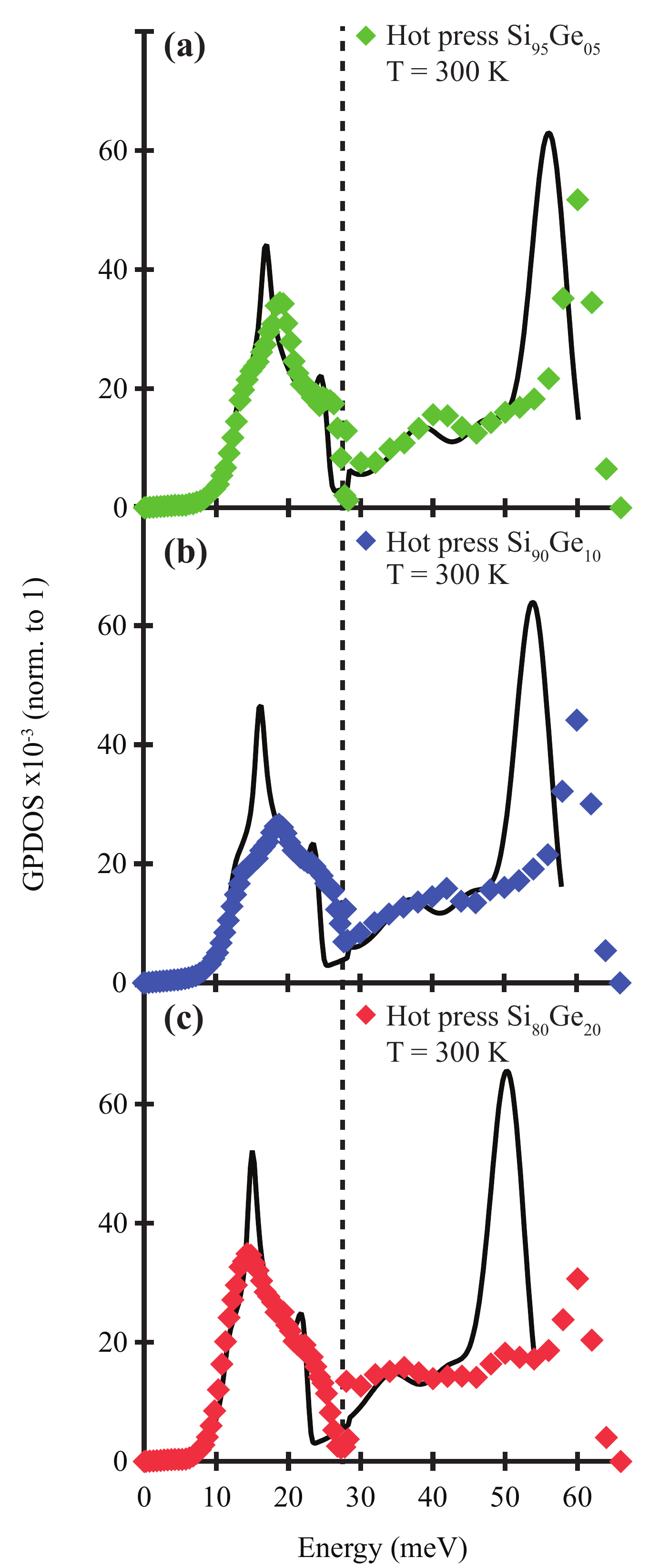}
\caption{Reduced 300 K GPDOS for the hot pressed (a) Si$_{0.95}$Ge$_{0.05}$, (b) Si$_{0.90}$Ge$_{0.10}$, and (c) Si$_{0.80}$Ge$_{0.20}$ samples.   Solid lines are the corresponding DFT model calculations for the PDOS assuming the virtual crystal approximation.  Dashed line again denotes the crossover from $E_i=120$ meV data to $E_i=40$ meV data.}
\end{figure}

\section{Vibrational entropy change in nanopowder S\lowercase{i}}
The vibrational entropy change is defined\cite{fultz1995} as $\Delta S=-3k_B\int^{\infty}_{0}(PDOS_{nano} - PDOS_{bulk}) ln(E)\partial E$. In looking at the reduced PDOS for Si upon transitioning from the bulk to the nanopowder phase, a $\Delta S_{300K}=0.84\pm0.05$ $k_B/atom$ is calculated from the data at 300 K.  This unusually large entropy difference is likely an artifact from a slight over subtraction of the anomalous 300 K background scattering at low energies.  If we instead look at only the phonon spectra derived from the $E_i=120$ meV data above $\Delta E=8$ meV, then $\Delta S_{300K}=0.21\pm0.03$ $k_B/atom$ ---in line with values observed in a number of metallic nanocomposites \cite{fultz1995,fultzreview}.  Upon warming to 700 K, the anomalous background term in the nanocrystalline sample's scattering profile vanishes, and the $\Delta S_{700K}$ remains nearly unchanged at $0.20\pm0.03$ $k_B/atom$. After then cooling down and measuring at 500 K, the resulting $\Delta S_{500K}$ retains a similar value of $0.23\pm0.03$ $k_B/atom$.      

The differences in the lattice heat capacities between the bulk and nanopowder Si can also be calculated at each temperature through the relation: $C_v=3R\int^{\infty}_{0}PDOS(\omega)(\frac{\hbar\omega}{k_BT})^2\frac{e^{\hbar\omega/k_BT}}{(e^{\hbar\omega/k_BT}-1)^2}\partial\omega$ [$J$  $mole^{-1}$  $K^{-1}$].  As an initial check, the heat capacity of bulk Si at 300 K can be calculated using the measured PDOS giving $C_{v,bulk}=20.3\pm 0.4$.  This agrees reasonably well with known value of $C_{v}=19.9$  determined from calorimetry measurements \cite{glazov}. At 650 K, the heat capacity for bulk Si increases to $C_{v,bulk}=24.1\pm 0.5$ (near the expected value of $24.8$ \cite{glazov}) while for the nanocrystalline sample the heat capacity shows a slightly smaller value of $C_{v,nano}=23.9\pm 0.4$.  After cooling to 500 K, $C_{v,bulk}=23.5 \pm 0.4$ ($23.6$ expected \cite{glazov}) and $C_{v,nano}=23.1\pm 0.3$.  Naively, one would expect the heat capacity of the nanopowder Si to be higher than that of the bulk Si based on the nonneglible difference in the vibrational entropy between the two; however the slight decrease in $C_v$ arises due to the preferential weighting of higher frequency differences in the PDOS by the $ln(E)$ term in the $\Delta S$ calculation relative to the Einstein distribution weighting term in the $C_v$ integral. 

\section{GPDOS in nanostructured S\lowercase{i}$_{1-x}$G\lowercase{e}$_{x}$}
We now turn to the evolution of the GPDOS in hot pressed, nanostructured thermoelectric S\lowercase{i}$_{1-x}$G\lowercase{e}$_{x}$ alloys with $x=$ 0, 0.05, 0.10, 0.20.  The same data collection and reduction procedure previously described for bulk and nanopowder Si samples was employed on these hot pressed thermoelectric samples. The resulting momentum integrated total scattering plus the single- and multiphonon components for each sample at $T=300$ K are plotted in Fig. 6. Here each Ge-doped concentration is overplotted with the scattering components from the hot pressed Si sample as a reference for changes in the raw scattering profile.  Solid lines are again the composite fit to the total scattering, and the scaled multiphonon background fits the scattering beyond the single phonon cutoff well in all cases.  The multiphonon contribution changes only slightly upon adding Ge to the system, and the largest changes occur at low energies as the single phonon spectral weight is redshifted downward.  

The resulting, reduced GPDOS spectra for each Ge-doped sample are plotted in Fig. 7. Overplotted with each GPDOS spectrum are the results of DFT calculations of the PDOS for the corresponding Si$_{1-x}$Ge$_{x}$ alloy assuming a uniform Si and Ge distribution.  As Ge is progressively alloyed into the Si$_{1-x}$Ge$_{x}$ matrix, the spectral weight near the longitudinal acoustic zone boundary is enhanced between 30 and 40 meV.  At lower energies, there appears an overall broadening of the acoustic phonon mode distribution relative to the parent Si material for small Ge concentrations ($x=0.05$ and $x=0.10$) and an overall shift to lower frequencies for the $x=0.20$ sample.  The resulting acoustic mode distribution for the $x=0.20$ composition suggests a lifetime broadened form of the DFT theoretical expectation; however a simple convolution of the DFT modeled PDOS with a damped harmonic oscillator line shape fails to model the data.  

For $x\leq0.10$, the response of the GPDOS to the addition of Ge mimics a simple lifetime broadening term in the PDOS of Si with no appreciable frequency shift with the exception of a low energy shoulder that builds with increased Ge concentration.  We note here that the normalized presentation of both $E_i=40$ and $E_i=120$ meV data fails at the $28$ meV transition region for the x=0.20 sample due to appreciable spectral weight present induced at the crossover point; however the above analysis holds since relative changes are only ascertained within a single $E_i$ setting for a given sample.     

Now turning to the higher frequency optical phonon modes, the addition of Ge smoothly suppresses the spectral weight of the 60 meV zone boundary modes corresponding to Si-Si vibrations.  The characteristic energy of the optical Van Hove singularity is left unchanged however.  This implies substantial clustering effects in Ge-doped samples and a resulting two-mode behavior\cite{genzel, lockwood, wakabayashi, rat, beraud}, illustrated most clearly by the appearance of a second optical phonon peak in the GPDOS at 50 meV in the $x=0.20$ sample.  In all samples however, there remain substantial regions of Si clusters that retain their characteristic Si-Si 60 meV peak.

\begin{figure}
\includegraphics[scale=.5]{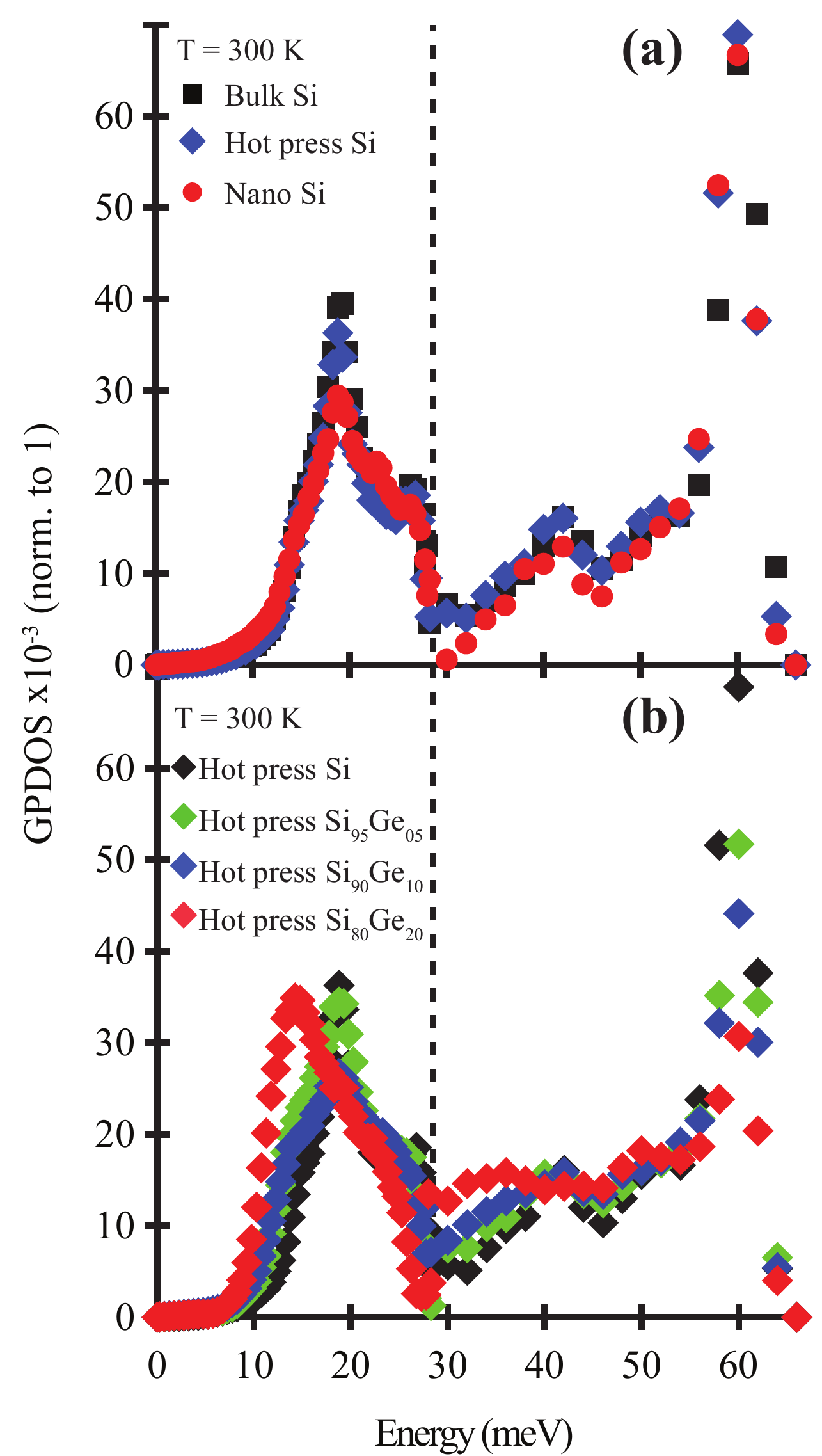}
\caption{(a) Comparison of the 300 K PDOS of the three pure Si samples in bulk (square symbols), hot pressed (diamond symbols), and nanopowder (circle symbols) forms. (b) Overplot showing the comparison of the 300 K GPDOS for the series of hot pressed thermoelectric Si$_{1-x}$Ge$_{x}$ samples with x=0, 0.05, 0.10, and 0.20.  Dashed line again denotes the crossover from $E_i=120$ meV data to $E_i=40$ meV data.}
\end{figure}

\section{Discussion}
A summarized comparison of the GPDOS determined for each sample measured at 300 K is show in Fig. 8.  In comparing the PDOS of the bulk, hot pressed, and nanopowder Si samples in Fig. 8 (a) there is no discernible shift in the spectral weight distribution of the PDOS between the bulk and hot pressed Si samples.  This stands in stark contrast to dramatic reduction in the thermal conductivity observed in the hot pressed Si sample\cite{renprl}, implying boundary modes and other scattering effects must take place at energies well below our experimental lower energy cutoff of $\Delta E=4$ meV or possess relative spectral weights below our experiments' sensitivity.  

In comparing the nanopowder Si sample to the bulk and hot pressed Si samples, there appears a substantial lifetime broadening effect resolvable in the acoustic regime which results in an enhanced tail in the PDOS extending down to the lowest energies measured. The effects of this lifetime broadening are removed once the nanopowder is hot pressed into a functional alloy; however the nanopowder PDOS spectrum may serve as a useful tool for amplifying otherwise subtle effects imparted via the high-energy ball milling process.

Now turning to the 300 K GPDOS data for the hot pressed Si$_{1-x}$Ge$_{x}$ alloys, a clear two-mode behavior appears as Ge is progressively doped into the material\cite{fenren}.  Phonon modes at the Si-Si transverse optical zone boundary retain their characteristic frequency while their relative spectral weight is damped with increasing Ge content.  This suppression in the spectral weight around $\approx 60$ meV is compensated by a broad redistribution/smearing of weight to frequencies as low as 30 meV.  At energies below 25 meV, the spectral weight distribution largely attributable to the transverse acoustic zone boundary modes in pure Si broadens with small Ge concentrations before undergoing a global shift to lower frequencies at $x=0.20$.  This implies that the clustering of Ge within the matrix at low concentrations primarily serves as a point defect scattering mechanism for reducing acoustic phonon lifetimes.  At large enough densities of Ge-impurities however, the lattice undergoes a global response and dispersion relations are renormalized.

The experimental picture of point defect scattering versus grain boundary scattering effects in Si$_{1-x}$Ge$_{x}$ therefore suggests that the Ge point defect scatterers, even at concentrations as low as 5$\%$, introduce a substantial change to the global GPDOS relative to simply hot pressing a frozen network of grain boundaries in pure Si.  At low enough Ge concentrations, the overall effect on the GPDOS at acoustic phonon frequencies empirically mimics that observed in pure nanopowder Si; however the potential thermal activation of grain boundary modes or defects milled into the Si nanocrystalline matrix at high temperatures and frequencies remains an issue for future experiments.

\section{Conclusions}
To conclude, we have investigated the evolution of the GPDOS in a series of nanostructured Si$_{1-x}$Ge$_{x}$ thermoelectric alloys as well as changes to the PDOS within nanopowder Si created via high-energy ball milling.  Ge-substitution into Si$_{1-x}$Ge$_{x}$ nanostructured alloys results in a two-mode type of behavior where clustering effects are evident.  For light Ge-concentrations with $x\leq0.10$, the acoustic phonon mode distributions broaden consistent with a substantial reduction in the acoustic phonon lifetime; however hot pressed, nanostructured Si exhibits a nearly identical PDOS to that of bulk Si despite an order of magnitude reduction in thermal conductivity.  This implies the dominant role of long wavelength and rare-region scattering mechanisms beyond the resolution of our experiments.  Our experiments do however resolve a dramatic lifetime broadening of acoustic phonon modes in pure nanopowder Si relative to the bulk along with an increase in the vibrational entropy of $\Delta S\approx0.20\pm0.03$ $k_B/atom$.  A temperature dependence was reported in the PDOS of nanopowder Si distinct from the bulk system that likely implies the thermal activation of additional modes within the nanocrystallite lattice.

\acknowledgments{
S.D.W. acknowledges helpful discussions with O. Delaire.  The work at BC was supported by NSF Awards DMR-1056625 (S.D.W) and CBET-1066634 (D.A.B) and also by the US Department of Energy under Contract Number DOE DE-FG02-00ER45805 (Z.F.R.).  Part of this work was performed at ORNL's SNS, sponsored by the Scientific User Facilities Division, Office of Basic Energy Sciences, U.S. DOE.}


\begin{thebibliography}{}
\bibitem{poudel} B. Poudel, Q. Hao, Y. Ma, Y. Lan, A. Minnich, B. Yu, X. Yan, D. Wang, A. Muto, D. Vashaee, X. Chen, J. Liu, M. Dresselhaus, G. Chen, and Z. F. Ren, Science 320, 634 (2008).
\bibitem{yu}  Bo Yu, Mona Zebarjadi, Hui Wang, Kevin Lukas, Hengzhi Wang, Dezhi Wang, Cyril Opeil, Mildred Dresselhaus, Gang Chen, and Zhifeng Ren, Nano Lett. 12, 2077 (2012).
\bibitem{luoptics} J. Lu et al., J. Alloys Compd. 341, 220 (2002).
\bibitem{yang} R.G. Yang and G. Chen, Phys. Rev. B 69, 195316 (2004).
\bibitem{hsu} K. F. Hsu, S. Loo, F. Guo, W. Chen, J.S. Dyck, C. Uher, T. Hogan, E. K. Polychroniadis, and M.G. Kanatzidis, Science, 303, 818 (2004).
\bibitem{zhao} X. B. Zhao, X. H. Ji, Y. H. Zhang, T. J. Zhu, J. P. Tu, and X. B. Zhang, Appl. Phys. Lett. 86, 062111 (2005).
\bibitem{dhital} Chetan Dhital, Clarina de la Cruz, C. Opeil, A. Treat, K.F. Wang, J.-M. Liu, Z.F. Ren, and Stephen D. Wilson, Phys. Rev. B. 84, 144401 (2011).
\bibitem{renprl} G. H. Zhu, H. Lee, Y. C. Lan, X.W. Wang, G. Joshi, D. Z. Wang, J. Yang, D. Vashaee, H. Guilbert, A. Pillitteri, M. S. Dresselhaus, G. Chen, and Z. F. Ren, Phys. Rev. Lett. 102, 196803 (2009). 
\bibitem{ma} Yi Ma, Qing Hao, Bed Poudel, Yucheng Lan, Bo Yu, Dezhi Wang, Gang Chen, and Zhifeng Ren, Nano. Lett. 8, 2580 (2008).
\bibitem{dresselhouse} M. S. Dresselhaus, G. Chen, Z. F. Ren, J. P. Fleurial, and P. Gogna, Adv. Mater. 19, 1043 (2007).
\bibitem{sopu} D. Sopu, J. Kotakoski, and K. Albe, Phys. Rev. B 83, 245416 (2011).
\bibitem{fultzphilmag} H. N. Frase, L. J. Nagel, J. L. Robertson, and  B. Fultz, Phil. Mag. B 75, 335 (1997).
\bibitem{yangraman} C. C. Yang and S. Li, J. Phys. Chem. B, 112 14193 (2008).
\bibitem{gupta} Sanjeev K. Gupta and Prafulla K. Jha, Sol. Stat. Comm. 149, 1989 (2009).
\bibitem{wangraman} Yu Wang, Lin Dong, Xiaolin Jia, Deliang Chen, and Shaokang Guan, J. Nanosc. and Nanotec. 11, 3592 (2011).
\bibitem{kara} Abdelkader Kara and Talat S. Rahman, Phys. Rev. Lett. 81, 1453 (1998).
\bibitem{fultzreview} For a review:  B. Fultz, Prog. Mater. Sci. 55, 247 (2010).
\bibitem{fultzjap} B. Fultz, J. L. Robertson, T. A. Stephens and L. J. Nagel, and S. Spooner, J. Appl. Phys. 79, 11 (1996).
\bibitem{frase} H. Frase, B. Fultz, and J. L. Robertson, Phys. Rev. B 57, 898 (1998).
\bibitem{fenren} Shang-Fen Ren, Wei Cheng, and Peter Y. Yu, Phys. Rev. B 69, 235327 (2004).
\bibitem{baroni} S. Baroni et al., http://www.quantum-espresso.org.
\bibitem{monkhorst} H.J. Monkhorst and J.D. Pack, Phys. Rev. B 13, 5188 (1976).
\bibitem{barth} U. von Barth and R. Car (unpublished); for a brief description of this method, see A. Dal
Corso, S. Baroni, R. Resta and S. de Gironcoli, Phys. Rev. B 47, 3588 (1993).
\bibitem{ceperly} D.M. Ceperley and B.J. Alder, Phys. Rev. Lett. 45, 566 (1980).
\bibitem{perdew} J.P. Perdew and A. Zunger, Phys. Rev. B 23, 5048 (1981).
\bibitem{broido} D. A. Broido, M. Malorny, G. Birner, N. Mingo and D. A. Stewart, Appl. Phys. Lett. 91, 231922 (2007).
\bibitem{ward}	A. Ward, D. A. Broido, D. A. Stewart and G. Deinzer, Phys. Rev. B 80, 125203 (2009).
\bibitem{kundu}	A. Kundu, N. Mingo, D. A. Broido, and D. A. Stewart, Phys. Rev. B 84, 125426 (2011).
\bibitem{abernathy} D. L. Abernathy, M. B. Stone, M. J. Loguillo, M. S. Lucas, O. Delaire, X. Tang, J. Y. Y. Lin, and  B. Fultz, Rev. Sci. Instr. 83, 15114 (2012).
\bibitem{kresch} M. Kresch, O. Delaire, R. Stevens, J. Y. Y. Lin, and B. Fultz, Phys. Rev. B 75, 104301 (2007).
\bibitem{williamson} G. K. Williamson and W. H. Hall, Acta Metallurgica 1, 22 (1953).
\bibitem{nilsson} G. Nilsson and G. Nelin, Phys. Rev. B 6, 3777 (1972).
\bibitem{nelin} G. Nelin and G. Nilsson, Phys. Rev. B 5, 3151 (1971).
\bibitem{glazov} V. M. Glazov and A. S. Pashinkin, High Temperature 39, 414 (2001).
\bibitem{fultz1995} B. Fultz, L. Anthony, L. J. Nagel, R. M. Nicklow, and S. Spooner, Phys. Rev. B 52, 3315 (1995).
\bibitem{delaire} O. Delaire, A. F. May, M. A. McGuire, W. D. Porter, M. S. Lucas, M. B. Stone, D. L. Abernathy, V. A. Ravi, S. A. Firdosy, and G. J. Snyder, Phys. Rev. B 80, 184302 (2009).
\bibitem{fultz1997} B. Fultz, C. C. Ahn, E. E. Alp, W. Sturhahn, and T. S. Toellner, Phys. Rev. Lett. 79, 937 (1997).
\bibitem{genzel} L. Genzel, T. P. Martin, and C. H. Perr, Phys. Stat. Sol. (B) 62, 83 (1974).
\bibitem{lockwood} D. J. Lockwood, K. Rajan, E. W. Fenton, J.-M. Baribeaue, and M. W. Denhoff, Solid State Commun. 61, 465 (1987).
\bibitem{wakabayashi} N. Wakabayashi, R. M. Nicklow, and H. G. Smith, Phys. Rev. B 4, 2558–2560 (1971).
\bibitem{rat} E. Rat, B. Hehlen, J. Kulda, I. Yonenaga, H. Casalta, E. Courtens, M. Foret, and R. Vacher, Physica B 276, 429 (2000)
\bibitem{beraud} A Béraud, J Kulda, I Yonenaga, M Foret, B Salce, and E Courtens, Physica B 350, 254 (2004).
\end{thebibliography}

\end{document}